\documentclass[twocolumn]{aastex63}

\usepackage{amsmath,amssymb,lineno}
\usepackage{newtxmath}
\usepackage{cleveref}
\usepackage{nicefrac}
\usepackage{upgreek}
\usepackage{todonotes}

\submitjournal{ApJ}

\shorttitle{}
\shortauthors{}

\newcommand{\dif}{\mathrm{d}}

\begin{document}

\title{Floor of cosmogenic neutrino fluxes above $10^{17}~$eV}

\author[0000-0001-8294-6294]{Corinne B\'erat}
\affiliation{Laboratoire de Physique Subatomique \& Cosmologie,
CNRS/IN2P3, Universit\'{e} Grenoble Alpes, Grenoble, France}

\author[0000-0001-6841-3280]{Antonio Condorelli}
\affiliation{Laboratoire de Physique des 2 Infinis Ir\`ene Joliot-Curie,
CNRS/IN2P3, Universit\'{e} Paris-Saclay, Orsay, France}

\author[0000-0001-6863-6572]{Olivier Deligny}
\affiliation{Laboratoire de Physique des 2 Infinis Ir\`ene Joliot-Curie,
CNRS/IN2P3, Universit\'{e} Paris-Saclay, Orsay, France}

\author[0000-0001-9787-596X]{Fran\c{c}ois Montanet}
\affiliation{Laboratoire de Physique Subatomique \& Cosmologie,
CNRS/IN2P3, Universit\'{e} Grenoble Alpes, Grenoble, France}

\author[0000-0002-8327-8459]{Zo\'e Torr\`es}
\affiliation{Laboratoire de Physique Subatomique \& Cosmologie,
CNRS/IN2P3, Universit\'{e} Grenoble Alpes, Grenoble, France}

\begin{abstract}

The search for neutrinos with energies greater than $10^{17}~$eV is being actively pursued. Although normalization of the dominant neutrino flux is highly uncertain, a floor level is guaranteed by the interactions of extragalactic cosmic rays with Milky Way gas. We estimate that this floor level gives an energy flux of $E^2\phi_\nu\simeq 10^{-13^{+0.5}_{-0.5}}~$GeV~cm$^{-2}$~sr$^{-1}$~s$^{-1}$ at $10^{18}~$eV, where uncertainties arise from the modeling of the gas distribution and the experimental determination of the mass composition of ultra-high-energy cosmic rays on Earth. Based on a minimal model of cosmic-ray production to explain the mass-discriminated energy spectra observed on Earth above $5{\times}10^{18}$~eV, we also present generic estimates of the neutrino fluxes expected from extragalactic production that generally exceed the aforementioned guaranteed floor. The prospects for detecting neutrinos above $10^{18}$~eV remain however challenging, unless proton acceleration to the highest energies is at play in a sub-dominant population of cosmic-ray sources or new physical phenomena are at work.

\end{abstract}



\section{Introduction} 
\label{sec:intro}

The energy spectrum of cosmic rays has long been observed to extend beyond $10^{20}~$eV since the first evidence for a primary particle with such an energy~\citep{Linsley:1963km}. Charged particles are thus accelerated to such ultra-high energies (UHE) in powerful astrophysical objects, the identification of which is still actively pursued. As a result of their interactions in the environment of the sources or \textit{en route} to Earth, neutrinos are produced with an energy corresponding to a fraction of the energy of cosmic rays. Detection of these cosmogenic neutrinos at energies above $10^{17}$~eV is a major challenge for astroparticle observatories.

UHE neutrino fluxes guaranteed by the interactions of cosmic rays propagating to Earth with the background photon fields permeating the Universe, most notably the cosmic microwave background, have long been considered in the literature~\citep[e.g.][]{Hill:1983mk,Protheroe:1995ft,Lee:1996fp,Waxman:1998yy,Engel:2001hd,Ahlers:2010fw,Kampert:2011hkm,PierreAuger:2022atd,PierreAuger:2023mid}. However, their precise knowledge relies on assumptions that can change the expectations by orders of magnitude. The main production channel is the decay of $\pi^\pm$ mesons. The hadrons that cause the creation of these mesons may be primary proton cosmic rays, or secondary mainly produced by the photo-disintegration of nuclei interacting inelastically with a cosmic background photon. Since the nucleons produced in a photo-disintegration inherit the energy of the fragmented nucleus divided by its atomic number, the neutrinos ultimately produced from primary heavy nuclei are of lower energies than those from lighter ones or from proton primaries. The neutrino flux, therefore, depends primarily on the cosmic-ray mass composition, which remains poorly constrained above about $5\times 10^{19}$~eV. Other important dependencies come from the maximum acceleration energy of the cosmic rays at the sources, the shape of the energy spectrum of the accelerated particles, and the cosmological evolution of the sources. As a consequence of the various progresses made over time to constrain these quantities, flux predictions at $10^{18}$~eV went from fairly high values, namely energy fluxes up to $E^2\phi_\nu\simeq 10^{-9}~$GeV~cm$^{-2}$~sr$^{-1}$~s$^{-1}$, obtained for a vanilla pure-proton composition to much lower ones, namely $E^2\phi_\nu\simeq 10^{-12}~$GeV~cm$^{-2}$~sr$^{-1}$~s$^{-1}$, in the framework of a mixed-composition model much more inline with the various constraints inferred from the data collected at the Pierre Auger Observatory~\citep{PierreAuger:2014gko,PierreAuger:2014sui,PierreAuger:2023xfc} and in other experiments~\citep{Watson:2021rfb}.  

UHE neutrinos are also expected from interactions of cosmic rays in their source environments. Such interactions have proved to be a key input for shaping the energy spectra of particles ejected from the sources~\citep{Globus:2015xga,Unger:2015laa,Biehl:2017zlw,Fang:2017zjf,Supanitsky:2018jje}. The counterpart in neutrino energy fluxes of such numerous interactions can be larger by several orders of magnitude than $E^2\phi_\nu\simeq 10^{-12}~$GeV~cm$^{-2}$~sr$^{-1}$~s$^{-1}$ at $10^{18}$~eV and even flirts with the current sensitivities of the Ice Cube and Pierre Auger observatories~\citep{Biehl:2017hnb,Boncioli:2018lrv,Muzio:2022bak,Condorelli:2022vfa}. It suffers however from additional uncertainties to those already aforementioned, such as the content in gas and photon fields of the source environments, and the confinement time of the particles.

Last but not least, cosmogenic neutrinos are also produced in the interstellar matter of the Galactic disk irradiated by UHE cosmic rays, in the same way as those of lower energy recently reported in~\cite{IceCube:2023ame} produced by Galactic cosmic rays. The flux expected from these interactions has received little attention in the literature. In this paper, therefore, we aim at estimating the contribution of UHE cosmic ray interactions in the Galaxy to the cosmogenic neutrino fluxes above $10^{17}~$eV. In contrast to the contributions mentioned above, the calculation of this flux does not resort to modeling assumptions that make the estimate spread over orders of magnitude. It suffers only from uncertainties in the mass composition observed on Earth and in the gas density in the Galaxy. The estimate can therefore be considered as a guaranteed floor of cosmogenic neutrinos.  

The paper is organized as follows. In Section~\ref{sec:UHEnuMW}, we describe the mass-discriminated energy spectra of cosmic rays on Earth, a survey of modelings of the interstellar gas density in the disk of the Milky Way, and the modelings of the neutrino production through the interactions of interest. The resulting neutrino flux is calculated in Section~\ref{sec:floor}; particular care is given to the estimation of the related systematics uncertainties. Neutrino fluxes produced outside the Galaxy from interactions in source environments and propagation effects are estimated in Section~\ref{sec:extragalactic}, assuming the cosmic ray spectra to be shaped by the source environments. Finally, the significance of the results is discussed in Section~\ref{sec:discussion}.

\section{UHE Neutrino production in the Milky Way} 
\label{sec:UHEnuMW}

\subsection{UHE cosmic ray benchmarks} 
\label{sec:UHECRs}

UHE cosmic rays are detected through indirect observations, using the extensive air showers they cause in the atmosphere. Under these conditions, only a statistical analysis can be used to determine the mass-dependent energy spectrum $\phi_i(E)$ of each nuclear component, by combining the all-particle energy spectrum with the energy-dependent abundances of elements. To estimate $\phi_i(E)$, we use the energy spectrum of the Pierre Auger Observatory reported above $10^{17}$~eV in \cite{PierreAuger:2021hun}, weighted by the fraction of elements separated in four mass groups (protons p, helium He, carbon-nitrogen-oxygen CNO, iron Fe) reported in \cite{Bellido:2017cgf} and \cite{PierreAuger:2023xfc}. This Observatory is currently providing the largest cumulative exposure with a single detector type, avoiding the need to combine measurements that inevitably introduces additional systematic effects. 

\begin{figure}[t]
\centering
\includegraphics[width=\columnwidth]{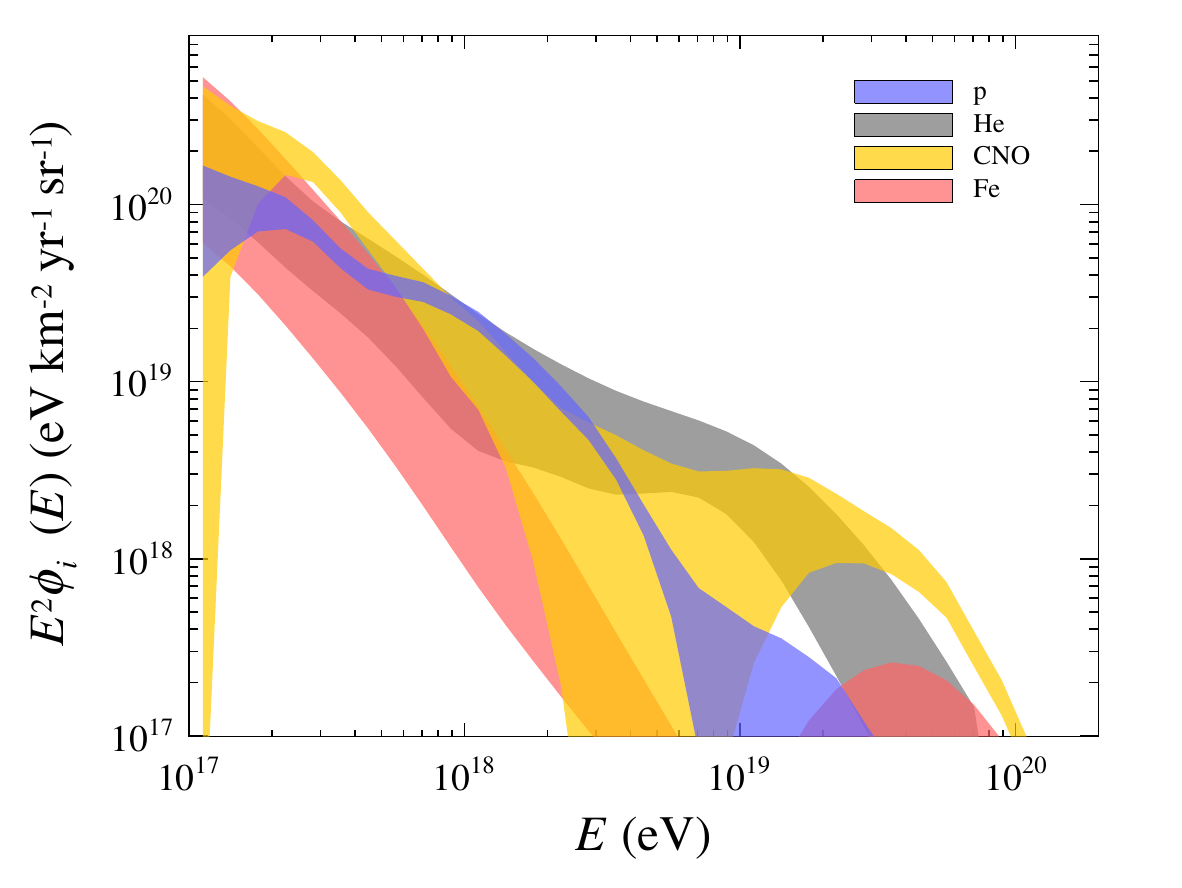}
\caption{Mass-discriminated energy-flux spectra of UHE cosmic rays inferred from data of the Pierre Auger Observatory~\citep{PierreAuger:2021hun,Bellido:2017cgf,PierreAuger:2023xfc}.}
\label{fig:E2fluxes}
\end{figure}

The mass-discriminated energy-flux spectra of UHE cosmic rays are shown in Fig.~\ref{fig:E2fluxes}, where a linear interpolation (in decimal logarithm) is applied to smooth the fluxes at the bin-center values. Two hadronic-interaction generators are used to model the development of the showers and to infer subsequently the mass composition, namely EPOS-LHC~\citep{Pierog:2013ria} and  Sibyll2.3c~\citep{riehn2017hadronic}. They lead to energy-dependent shifts in the spectra as well as do the systematic uncertainties in the shower observables sensitive to mass composition. As a result, the current knowledge of the mass-discriminated energy spectra is limited to the color-coded bands. Despite the  uncertainties, the main features captured by these measurements can be summarized as follows. A component of Fe nuclei is observed to falloff steeply above $10^{17}~$eV, along the lines of the long-standing scenario for the upper end of Galactic cosmic rays characterised by a rigidity-dependent maximum acceleration energy~\citep{1961NCim...22..800P}. On the other hand, the falloff of the component of protons, helium, and CNO-group nuclei at the highest energies, with a hint of recovery of Fe nuclei, is well reproduced by nuclear components that drop off at the same magnetic rigidity in extragalactic sources featuring a hard spectral index~\citep{Aloisio:2013hya,Taylor:2015rla,PierreAuger:2016use}. Such hard values for the spectral index could reflect the role of interactions in the source environments for shaping the ejected spectra, on the condition that the index of protons ejected from the sources is softer ~\citep{Globus:2015xga,Unger:2015laa,Biehl:2017zlw,Fang:2017zjf,Supanitsky:2018jje}. It was indeed shown that such a requirement is consistent with the data by considering the proton spectrum in the energy range across the ankle feature as an additional constraint~\citep{Luce:2022awd}. Furthermore, in the energy range between $10^{17}~$eV and $\simeq 4{\times}10^{17}~$eV ($\simeq 1{\times}10^{18}~$eV) [$\simeq 5{\times}10^{18}~$eV], another phenomena is called for producing the observed abundances of protons (helium) [CNO group]. Whether these elements are fueled by a Galactic event or an extragalactic one, we note that the ``beam'' of interest for the estimation of the neutrino flux produced in the Milky Way above $10^{17}~$eV is that of cosmic rays with energies larger than $\sim A{\times}10^{18}~$eV, with $A$ the atomic number of the particles. Hence, the origin of nuclei between $10^{17}$ and $10^{18}~$eV is not critical for estimating the neutrino fluxes sought for and we shall follow the assumption that all particles illuminate the Galaxy uniformly, as do those of extragalactic origin.

\subsection{Gas targets} 
\label{sec:gas}

Most of the mass in the interstellar medium in the Galaxy (a few $10^9~M_\sun$) is distributed predominantly in the disk and is made by hydrogen ($\simeq 90\%$) and helium ($\simeq 10\%$) in gaseous state. A comprehensive description of the data available to reconstruct the spatial distribution of the gas can be found in, e.g.,~\cite{Ferriere:2001rg}; only the main features necessary to the neutrino-flux calculation are reminded here. The component in atomic form \mbox{H\,{\sc i}}, which is probed using the line at 21 cm observed in emission or absorption, represents a large fraction of the mass. Its density is inferred to be constant in the distance range from 4 to $\simeq 10~$kpc from the Galactic center and to decrease steadily at larger distances. In the vertical direction in Galactic latitude, it falls exponentially with a scale length of $\simeq 250$~pc or $\simeq 130$~pc, depending whether the component is warm or cold. Almost equally abundant, the distribution of hydrogen in molecular form H$_2$ is less well known. It is observed in the ultraviolet region via electronic transitions and indirectly inferred from the observations at radio wavelengths of the $^{12}$C$^{16}$O molecule, which is excited by collisions with H$_2$. The surveys of the 21-cm line are thus supplemented by the integrated intensity of the $^{12}$C$^{16}$O lines. This intensity is almost linearly related to the column density of H$_2$ with a proportionality factor being taken as $2{\times}10^{20}~$cm$^{-2}~$(K~km~s$^{-1}$)$^{-1}~$\citep{Bolatto:2013ks}. The distribution of the helium contribution, accounting for around 10\%, is assumed to follow closely the hydrogen one. 

Following the same strategy as in \cite{Berat:2022iea}, we use two models of the spatial distribution of the gas and consider the differences in the final neutrino emission as contributing to the systematic uncertainties of  $\phi^{\mathrm{gal}}_\nu(E,\mathbf{n})$. The first model, developed by~\cite{Lipari:2018gzn} and dubbed as model A hereafter, does not strive to correctly describe minute features; it aims at capturing the large-scale properties of the gas based on an axially and up-down symmetric distribution with a scale height increasing as a function of the radial distance from the Galactic center. The second model, developed by~\cite{Johannesson:2018bit} and dubbed as model B, incorporates in addition spiral arms and accounts for the warping of the disk.

\subsection{UHE cosmic ray-gas interactions} 
\label{sec:UHECR-gas}

The gas density rises to $\sim 1~$cm$^{-3}$ in some regions of the disk of the Milky Way. Such densities induce a low rate of cosmic ray-gas interactions that lead to the production of mesons including charged pions, which eventually produce neutrinos in their decay byproducts. The inelastic cross section for a cosmic-ray element $i$ with energy $E'$ and a gas element $j$ at rest, $\sigma_{ij}(E')$, is extracted using cosmic-ray event generators such as EPOS-LHC or Sibyll2.3 emulated in the Cosmic Ray Monte Carlo (CRMC) package \citep{crmcsource}. It ranges typically in the hundred of millibarns, reaching thousand ones for the heaviest collisions. 

\begin{figure}[t]
\centering
\includegraphics[width=\columnwidth]{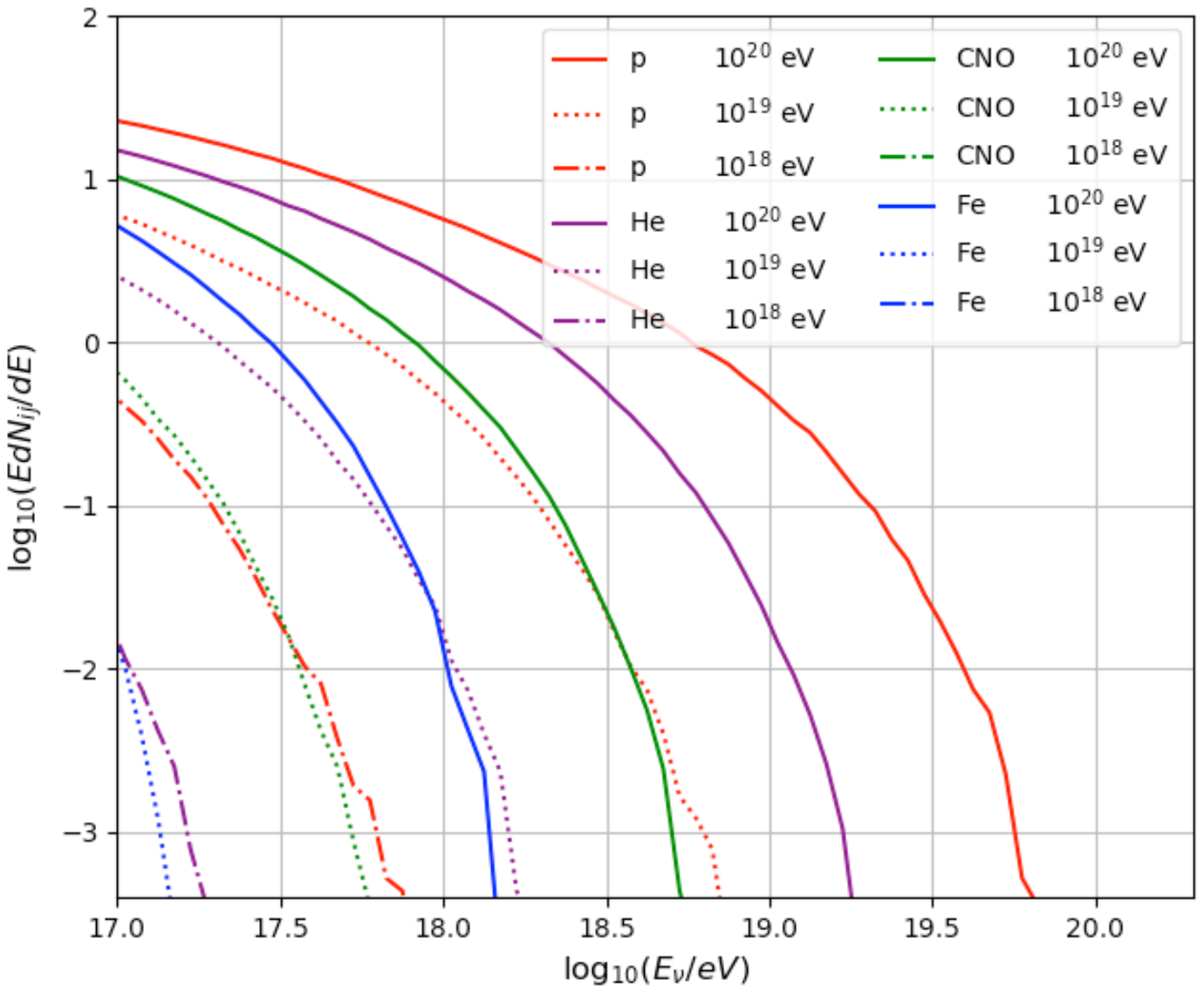}
\caption{Neutrino yield per cosmic ray interaction for different primaries and incoming energies.}
\label{fig:yield}
\end{figure}

Correspondingly, the inclusive spectrum of neutrinos $\dif N^{\nu}_{ij}/\dif E(E',E)$, which corresponds to the mean number of neutrinos in the energy range $[E,E+\dif E]$ produced in a single interaction, is obtained through the follow-up provided by CRMC of secondary particle decays that are expected to result in neutrinos. For the various primaries and three cosmic ray energies, $10^{18}~$eV (dotted), $10^{19}~$eV (dashed) and $10^{20}~$eV (continuous), the neutrino yields are displayed in Fig.~\ref{fig:yield}. The expected increase in yield with incident cosmic-ray energy is observed for a fixed neutrino energy $E<E'$. On the other hand, it is also noted that the yield substantially depends on the cosmic-ray mass for a fixed cosmic-ray energy. Since only one nucleon typically participates in each interaction, this behavior is predicted by the energy present in each nucleus, which is diminished by the atomic number $A$ when compared to the overall energy of the nucleus. As a result, neutrino fluxes are produced in greater quantities by lightest primaries at higher energy. 

Neutrinos of each flavor are not produced equally through the decay of the charged mesons. However, given the Pontecorvo-Maki-Nakagawa-Sakata matrix, it is reasonable to expect complete mixing of flavors from oscillations over propagation distances of the order of kiloparsecs, and thus an equal flux for each flavor on Earth.

\section{Floor of UHE-neutrino fluxes} 
\label{sec:floor}

The diffuse neutrino flux (per steradian) at energy $E$ produced from cosmic ray-gas interactions in the Galaxy can be estimated in the thin-target regime by integrating the position-dependent emission rate per unit volume and unit energy along the line of sight $s$,
\begin{equation}
    \label{eqn:nuflux}
    \phi^{\mathrm{gal}}_\nu(E,\mathbf{n})=\frac{1}{4\pi}\int_0^\infty \dif s~q^{\mathrm{gal}}_\nu(E,\mathbf{x}_\sun+s\mathbf{n}).
\end{equation}
Here, $\mathbf{x}_\sun$ is the position of the Solar system in the Galaxy and $\mathbf{n}\equiv\mathbf{n}(\ell,b)$ is a unit vector on the sphere pointing to the longitude $\ell$ and latitude $b$, in Galactic coordinates. The $1/4\pi$ factor models the isotropic emission from anywhere in the Galaxy due to the isotropic irradiation of the gas by extragalactic UHE cosmic rays. The neutrino emission stems from the creation and decay of unstable mesons and subsequent leptons in the inelastic interactions of cosmic rays with the different interstellar-gas elements $j$ with density $n_j(\mathbf{x})$,
\begin{multline}
    \label{eqn:nusource}
    q^{\mathrm{gal}}_\nu(E,\mathbf{x})=\\
    4\pi\sum_{i,j}n_j(\mathbf{x})\int_E^\infty\dif E'\phi_i(E')\sigma_{ij}(E')\frac{\dif N_{ij}^\nu}{\dif E}(E',E).
\end{multline}
The integration is carried out over all cosmic ray energies $E'>E$ that allow for generating neutrinos with energy $E$. The $4\pi$ factor results from the integration of the cosmic ray flux, considered isotropic, over solid angle.

\begin{figure}[t]
\centering
\includegraphics[width=\columnwidth]{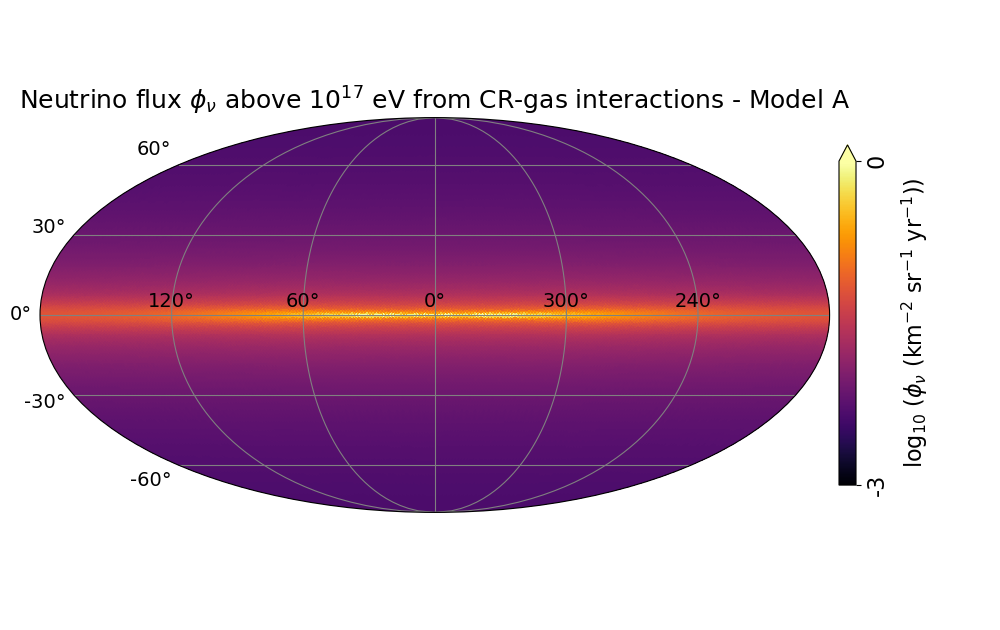}
\caption{Neutrino flux in Galactic coordinates (Hammer
projection) expected from cosmic ray-gas interactions in the Milky Way, integrated above $10^{17}$~eV. Model A is used for the gas distribution in the Galactic disk.}
\label{fig:skygal-A}
\end{figure}
\begin{figure}[t]
\centering
\includegraphics[width=\columnwidth]{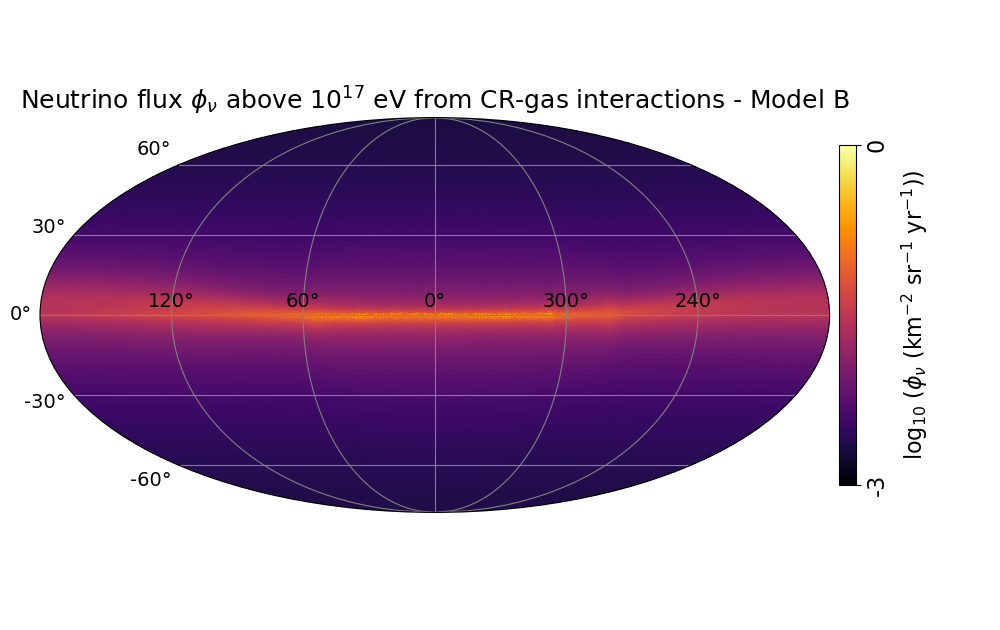}
\caption{Same as Fig.~\ref{fig:skygal-A}, using model B for the gas distribution in the Galactic disk.}
\label{fig:skygal-B}
\end{figure}

\begin{figure}[t]
\centering
\includegraphics[width=\columnwidth]{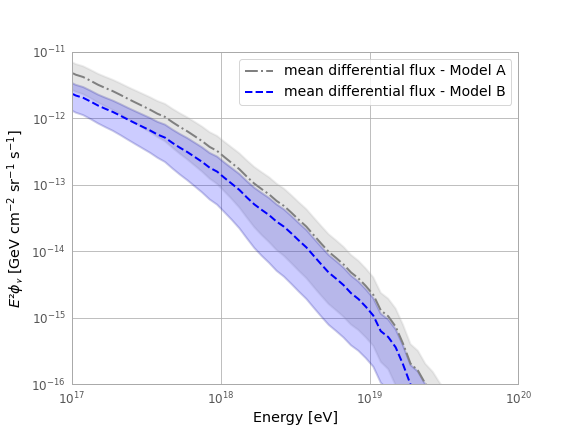}
\caption{Energy flux of neutrinos expected from cosmic ray interactions in the disk of the Milky Way for two models of the dust distribution.}
\label{fig:gal-neutrinos}
\end{figure}

The neutrino flux integrated above $10^{17}~$eV is shown as a function of the incoming direction in Fig.~\ref{fig:skygal-A} and Fig.~\ref{fig:skygal-B} for the two models of gas distribution in the Galactic disk. In each case, the benchmark mass-composition of cosmic rays is that inferred from the EPOS-LHC hadronic interaction generator. As expected, the flux is concentrated a few degrees around the Galactic plane, amounting to $\simeq 10{\times}10^{-2}$~km$^{-2}$yr$^{-1}$sr$^{-1}$ once averaged out over $|b|\leq5^\circ$ in the case of model A ($\simeq 4{\times}10^{-2}$~km$^{-2}$yr$^{-1}$sr$^{-1}$ in the case of model B). The pattern is brighter in the innermost region of the disk for both models. 

The energy flux averaged over full sky is shown in Fig.~\ref{fig:gal-neutrinos} as a function of energy. The energy dependence is shaped by that of the cosmic-ray flux, inheriting from its main features but shifted about a decade earlier. The bands correspond to the systematic uncertainties arising from those in the mass-discriminated energy spectra. These uncertainties are sourced by the systematics in determining the abundance of elements discussed in Section~\ref{sec:UHECRs} and those in determining the all-particle energy spectrum. To cope with these effects, realizations of the mass-discriminated energy spectra are constructed as follows. The hadronic interaction generator is selected randomly, and the abundance of elements is picked up within the corresponding systematics. The energy spectra for each element require, in addition to these abundances, a realization of the all-particle energy spectrum within its own systematics. Denoting as a vector the set of measurements in each energy bin, $\boldsymbol{\phi}=\{\phi_1,\phi_2,...,\phi_N\}$, the $\boldsymbol{\phi}_+$ ($\boldsymbol{\phi}_-$) vector is defined as the set of values that satisfy
\begin{eqnarray}
\label{eqn:quantile}
    \frac{1}{\sqrt{(2\pi)^N\mathrm{det}\boldsymbol{\sigma_\phi}}}\int_{{\boldsymbol{\phi}_+}}^{+\infty}\dif\boldsymbol{\phi}\exp{\left(-\frac{1}{2}\delta\boldsymbol{\phi}^\mathrm{T}\boldsymbol{\sigma^{-1}_\phi}\delta\boldsymbol{\phi}\right)}&=&C_+,\nonumber \\
    \frac{1}{\sqrt{(2\pi)^N\mathrm{det}\boldsymbol{\sigma_\phi}}}\int_{-\infty}^{\boldsymbol{\phi}_-}\dif\boldsymbol{\phi}\exp{\left(-\frac{1}{2}\delta\boldsymbol{\phi}^\mathrm{T}\boldsymbol{\sigma^{-1}_\phi}\delta\boldsymbol{\phi}\right)}&=&C_-,
\end{eqnarray}
with $C_+=0.84$ $(C_-=0.16)$. The notation $\delta\boldsymbol{\phi}$ stands for a random fluctuation around the set of observed values, while the covariance matrix $\boldsymbol{\sigma_\phi}$ is taken from \cite{PierreAuger:2021hun}. To solve equation~\ref{eqn:quantile} for the unknown $\boldsymbol{\phi}_+ (\boldsymbol{\phi}_-)$, we build the probability distribution function of $\boldsymbol{\phi}$ by whitening the covariance matrix in the same way as in \cite{Berat:2022iea}. By repeating a large number of times the procedure, the 2-sided 16\% quantiles defining $\boldsymbol{\phi}_+$ and $\boldsymbol{\phi}_-$, and ultimately those of the mass-discriminated energy spectra, are finally estimated. Once propagated, these uncertainties impact by about one order of magnitude the neutrino flux. 

We thus find that the single-flavor neutrino energy flux sought for amounts to $\simeq 10^{-13}~$GeV~cm$^{-2}~$sr$^{-1}~$s$^{-1}$ at $10^{18}~$eV. Although guaranteed as a floor of neutrino flux, we discuss next other cosmogenic fluxes that may dominate and be revealed earlier.

\section{Extragalactic neutrino production} 
\label{sec:extragalactic}

To compare the guaranteed floor of neutrino fluxes obtained in Section~\ref{sec:floor} with other model-dependent expectations, we now turn to the neutrino production in the extragalactic space and in the environments of the sources of UHE cosmic rays. 

\subsection{Production \textit{en route}}
\label{sec:en-route}

The UHE neutrino flux that results from cosmic ray interactions with the photon baths in the universe can be estimated by integrating the generation rate per energy unit and per comoving volume unit of each species, $q_A(E)$, over lookback time, the role of which is played by redshift:
\begin{multline}
    \label{eqn:nu-xgalflux}
    \phi^\mathrm{xgal}_\nu(E)=\\
    \frac{c}{4\pi}
    \sum_A\iint\dif z\dif E'\left|\frac{\dif t}{\dif z}\right|S(z)q_A(E')\frac{\dif\eta_{A\nu}(E,E',z)}{\dif E}.
\end{multline}
Here, $S(z)$ stands for the redshift evolution of the production of cosmic rays assumed in the following to scale with that of the star-formation rate, $\eta_{A\nu}(E,E')$ is the fraction of particles with energy $E'$ and mass number $A$ that produce neutrinos with energy $E$, and the relationship between cosmic time and redshift follows from the concordance model used in cosmology, \mbox{$(\nicefrac{\dif t}{\dif z})^{-1}=-H_0(1+z)\sqrt{\Omega_{\mathrm{m}}(1+z)^3+\Omega_{\Lambda}}$} with $H_0=70\:$km\:s$^{-1}$\:Mpc$^{-1}$ the Hubble constant at present time, $\Omega_{\mathrm{m}}\simeq0.3$ the density of matter (baryonic and dark matter) and $\Omega_{\Lambda}\simeq0.7$ the dark-energy density. We use the \textit{SimProp} package \citep{Aloisio:2012wj}, a software dedicated to UHE cosmic ray propagation, to estimate $\nicefrac{\dif\eta_{A\nu}(E,E',z)}{\dif E}$. The generation rate $q_A(E)$ required at the escape from the source environments to fuel the extragalactic counterpart of the energy-flux spectra is taken from \cite{Luce:2022awd}.  The hardness of the emissivity of He, CNO, Si and Fe nuclei is necessary to reproduce the little mixture between elements observed in Fig.~\ref{fig:E2fluxes} above $10^{19}~$eV. By contrast, the spectral index describing the proton emissivity can get much softer values. This is along the lines of a copious production of nucleons of energy $\nicefrac{E}{A}$ subsequent to the spallation of nuclei with mass number $A$ before escaping the source environments. 

\begin{figure}[t]
\centering
\includegraphics[width=\columnwidth]{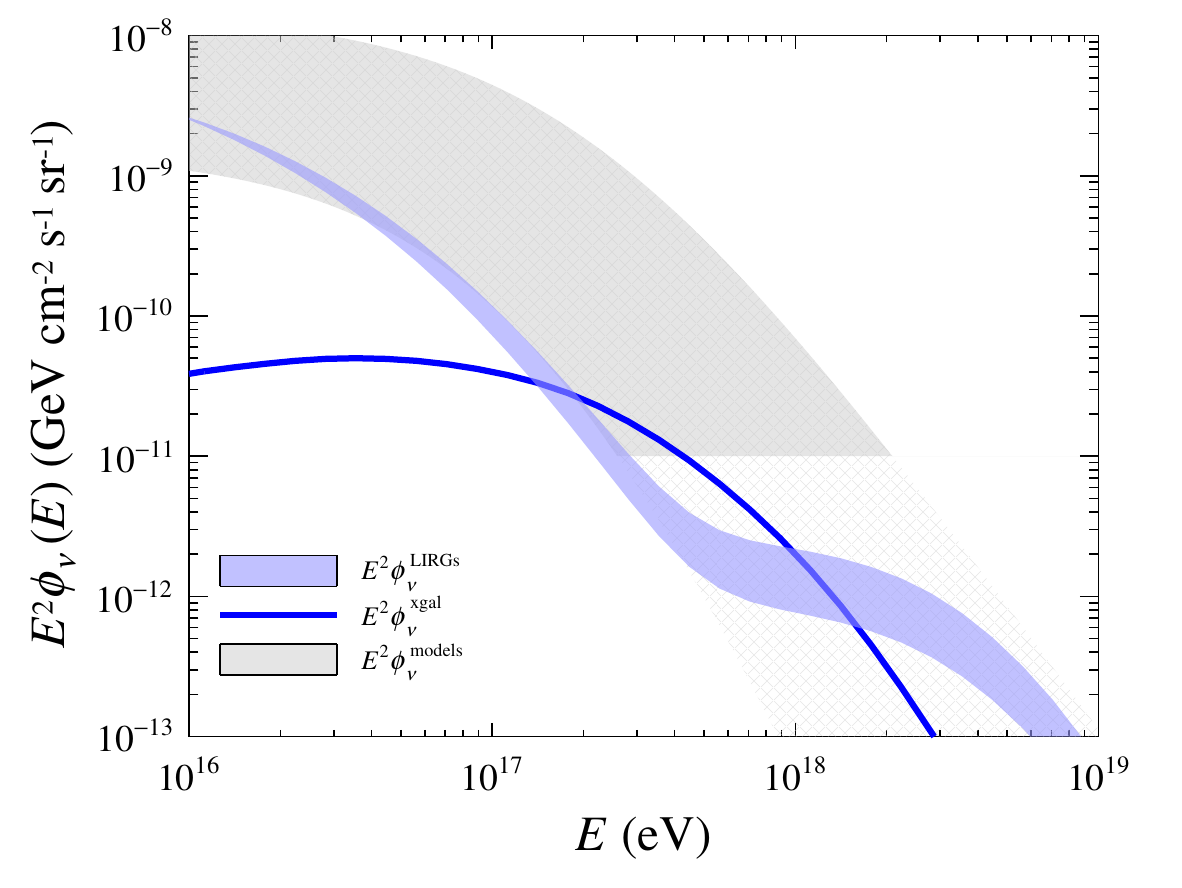}
\caption{Energy flux of neutrinos expected from  UHECR interactions in LIRGs (blue band) and from propagation in the extragalactic space (blue line). The grey band brackets various expectations from specific source models from \cite{Biehl:2017hnb,Zhang:2018agl,Boncioli:2018lrv,Muzio:2022bak,Condorelli:2022vfa} (the hatched area is an extrapolation of the results).}
\label{fig:xgal-neutrinos}
\end{figure}

The resulting energy flux is shown as the continuous blue line in Fig.~\ref{fig:xgal-neutrinos}, labeled as $E^2\phi^\mathrm{xgal}_\nu(E)$. As anticipated in the introduction, $E^2\phi^\mathrm{xgal}_\nu(E)$ peaks to a value smaller than $10^{-10}~$GeV~cm$^{-2}$~s$^{-1}$~sr$^{-1}$ below $10^{17}~$eV and rapidly falls off at higher energies, down to $10^{-12}~$GeV~cm$^{-2}$~s$^{-1}$~sr$^{-1}$ at $10^{18}~$eV. That $E^2\phi^\mathrm{xgal}_\nu(E)$ is significantly below the range anticipated at UHE in the early literature is primarily a consequence of the intermediate-to-heavy mass composition of cosmic rays.

\subsection{Production in the source environments}
\label{sec:in-source}

Ultimately, the generation rates $q_{A}(E)$ result from those injected by UHE cosmic ray sources, $\mathcal{Q}_{A}(E)$, and processed through propagation effects in source environments and/or hosting galaxies,
\begin{equation}
\label{eqn:emissivity}
    q_A(E)=\sum_{A'}\int_{\geq E}\dif E'\mathcal{Q}_{A'}(E')\frac{\dif\eta_{AA'}(E,E')}{\dif E},
\end{equation}
where $\eta_{AA'}(E,E')$ is the fraction of particles escaping the environments with energy $E$ and mass number $A$ from parent particles with energies $E'\geq E$ and mass numbers $A'\geq A$. We consider sources accelerating nuclei of charge $Z$ and mass number $A$ with a generation rate per energy unit and comoving volume unit conveniently parameterized as 
\begin{equation}
\label{eqn:fsupp}
    \mathcal{Q}_A(E) = \mathcal{Q}_{0A}\left(\frac{E}{E_0}\right)^{-\gamma}
    \begin{cases}
    1 & \mathrm{if~}E\leq E^{Z}_{\mathrm{max}},\\
    \exp{\left(1-\nicefrac{E}{E^{Z}_{\mathrm{max}}}\right)} & \mathrm{otherwise},
    \end{cases}
\end{equation}
with $\mathcal{Q}_{0A}$ the mass-dependent reference injection rates, $E^{Z}_{\mathrm{max}}=ZE_{\mathrm{max}}$ the maximum energy for each species, and $E_0$ an arbitrary pivot energy being taken as \mbox{$E_0=10^{18}$}~eV. 

As a generic source environment, we choose in the following luminous infrared galaxies, also known as LIRGs. These galaxies, quite abundant, host candidates of accelerators of UHE cosmic rays: active galactic nuclei as well as regions of high star formation rates that could be the origin of higher rates of gamma-ray bursts, relativistic supernovae or young neutron stars. The considered photon fields in LIRGs are two black bodies, one peaking in the optical range ($3.3{\times}10^{-1}~$eV) due to star light and another one in infrared ($3{\times}10^{-3}~$eV) due to ultraviolet reprocessing by the dusts \citep{Peretti:2018tmo}. The typical photon density is $\gtrsim 10^3 \, \rm eV \, cm^{-3}$ for either optical and infrared energy ranges, while the total infrared luminosity is $\simeq 10^{45}~\rm erg~s^{-1}$. The gas density is supposed to scale with that of the photons bath according to the \cite{Kennicutt_2012} relation. In parallel, a magnetic field of 1~$\upmu$G filling a sphere of $R\simeq 250$~pc-radius in the inner zone of the LIRGs is modelled as a Kolmogorov turbulence with a coherence length $l_\mathrm{coh}=1$~pc, as motivated by the typical scale at which the turbulence is expected to be injected in the nucleus of starburst galaxies \citep[see e.g.][]{Peretti:2018tmo}. 

For given generation rates per energy unit $\mathcal{Q}_A(E)$, the emissivities $E^2q_A(E)$ are calculated by means of simulations of test particles assuming a leaky-box model, as in \citep{Condorelli:2022vfa}: particles escape LIRGs if their interaction probability is smaller than their escape one; otherwise they loose energy and all their byproducts are accounted for in the following step of the simulation. The escape time is considered to be the minimum between the advection time, modelled as $t_{\rm adv}=R/v_{\rm W}$ with the wind speed $v_{\rm W} = 500 \ \rm km\ \rm s^{-1}$, and the diffusion one, modelled as $t_{\rm D}=R^2/D(E)$ with $D(E)$ the diffusion coefficient. The latter reads as $D(E) \simeq c r_{L}^{2-\delta}(E) \, l_\mathrm{coh}^{\delta -1} /3$, with $c$ the speed of light, $r_L$ the particle Larmor radius and $\delta$ the spectral slope of the turbulence, which for a Kolmogorov cascade reads as $\delta = 5/3$. Following \cite{Subedi:2016xwd}, we additionally consider the transition in the diffusion regime taking place when $r_L(E_\star) \sim l_\mathrm{coh}$, at which point the diffusion coefficient switches to $D(E) = D_\star(r_L(E)/l_\mathrm{coh})^2$, with $D_\star$ the value of the coefficient computed at the energy $E_\star$ such that $r_L(E_\star)= l_\mathrm{coh}$. A last transition occurs at high energy to guarantee that the diffusion time never goes below the free-escape time.

For the customary five representative mass elements $A=\{1,4,16,28,56\}$ and $E_{\mathrm{max}}=10^{18.4}~$eV, we find that the emissivities derived in \cite{Luce:2022awd} can be reproduced for reference generation rates scaling as $\mathcal{Q}_{0A}=\{0,0.020,0.800,0.179,0.001\}$ in units of $6{\times}10^{45}$~erg~Mpc$^{-3}$~yr$^{-1}$ and $\gamma=1.5$. The absence of injected protons is constrained by the energy scale of the corresponding component on Earth (Fig.~\ref{fig:E2fluxes}) observed to be equal to the energy per nucleon of the other components. This favors protons being fragments of primary nuclei, as already established in \cite{PierreAuger:2016use}. The fractions of Si and Fe nuclei, on the other hand, suffer from large and highly-correlated uncertainties as the plausible recovery of Fe at the highest energies is not yet established with existing data. Besides, we note that the value inferred for $\gamma$ can be accommodated with magnetic-reconnection acceleration mechanisms but not with first-order Fermi shock acceleration, in which case a benchmark is $\gamma=2$. However, additional in-source interactions for specific models of accelerators could be at play to contribute to shape the hardness of the spectra. In addition, little-known effects such as rigidity-dependent escape from magnetic fields within clusters of galaxies over Mpc scales~\citep{Donnert2018MagneticSimulation} or self-confinement of UHE cosmic rays through resonant streaming instabilities~\citep{Blasi:2019obb,Cermenati:2023gkm,Schroer:2023bfm} could also play a role. In the following, we therefore explore the dispersion of predictions for the neutrino fluxes by bracketing the range of $\gamma$ between 1.5 and 2. 

The neutrino flux that results from UHE cosmic rays interacting in LIRGs can be estimated by integrating the generation rate $q_\nu(E)$ over redshift: 
\begin{equation}
    \label{eqn:nu-LIRGflux}
    \phi^\mathrm{LIRGs}_\nu(E)=\frac{c}{4\pi}\int\dif z\left|\frac{\dif t}{\dif z}\right| S(z)q_\nu(E(1+z))e^{-\mathcal{O}_\nu(E,z)},
\end{equation}
where $\mathcal{O}_\nu(E,z)$ is the neutrino opacity of the universe at early times~\citep{Berezinsky:1991aa,Gondolo:1991rn}. Results are shown as the blue band in Fig.~\ref{fig:xgal-neutrinos}.   The higher expected fluxes result from the harder values of $\gamma$, as a result of the increased rate of interactions of cosmic rays of higher energy. A hardening is observed around $4{\times}10^{17}$~eV: the fluxes are primarily shaped by UHECR-gas interactions below this energy while they are governed by UHECR-photon interactions above. The fluxes expected from interactions during propagation in the extragalactic space are dominant by a factor of a few between $\sim 10^{17}$ and $\sim 10^{18}$~eV. 

Our estimate can be compared with those obtained in several studies that consider reproducing the mass-discriminated spectra above $5{\times}10^{18}~$eV in Fig.~\ref{fig:E2fluxes}, shown collectively as the grey band labeled as $E^2\phi_\nu^{\mathrm{models}}$ in Fig.~\ref{fig:xgal-neutrinos}. In \cite{Muzio:2022bak}, starburst-galaxy environments are shown to 
led to a rapid decrease of the energy flux in the range above $\simeq 10^{16.3}~$eV, energy at which $E^2\phi_\nu\simeq 10^{-8}~$GeV~cm$^{-2}$~sr$^{-1}$~s$^{-1}$. Similar results are obtained for in-source interactions in tidal disruption events~\citep{Biehl:2017hnb} or low-luminosity gamma ray bursts~\citep{Boncioli:2018lrv}: the energy flux, peaking just below $E^2\phi_\nu\simeq 10^{-8}~$GeV~cm$^{-2}$~sr$^{-1}$~s$^{-1}$ at $\simeq 10^{16}~$eV, falls off abruptly down to $E^2\phi_\nu\simeq 10^{-12}~$GeV~cm$^{-2}$~sr$^{-1}$~s$^{-1}$ at $\simeq 10^{18}~$eV. The largest expectations are those from  \cite{Zhang:2018agl} studying the case of engine-driven supernovae and from \cite{Condorelli:2022vfa} focusing on starburst galaxies. The increase value at $\simeq 10^{18}~$eV can, however, be attributed to an assumed lighter mass composition of cosmic rays at UHE than in other studies in the former case, or to an increased density of targets so as to reproduce the emission rates $q_A(E)$ from a single $\mathcal{Q}_{\mathrm{Si}}(E)$ generation rate in the latter one. Note that in all cases, we extrapolated the reported expectations below  $E^2\phi_\nu\simeq 10^{-11}~$GeV~cm$^{-2}$~sr$^{-1}$~s$^{-1}$ through the hatched band.
 
\section{Discussion} 
\label{sec:discussion}

\begin{figure}[t]
\centering
\includegraphics[width=\columnwidth]{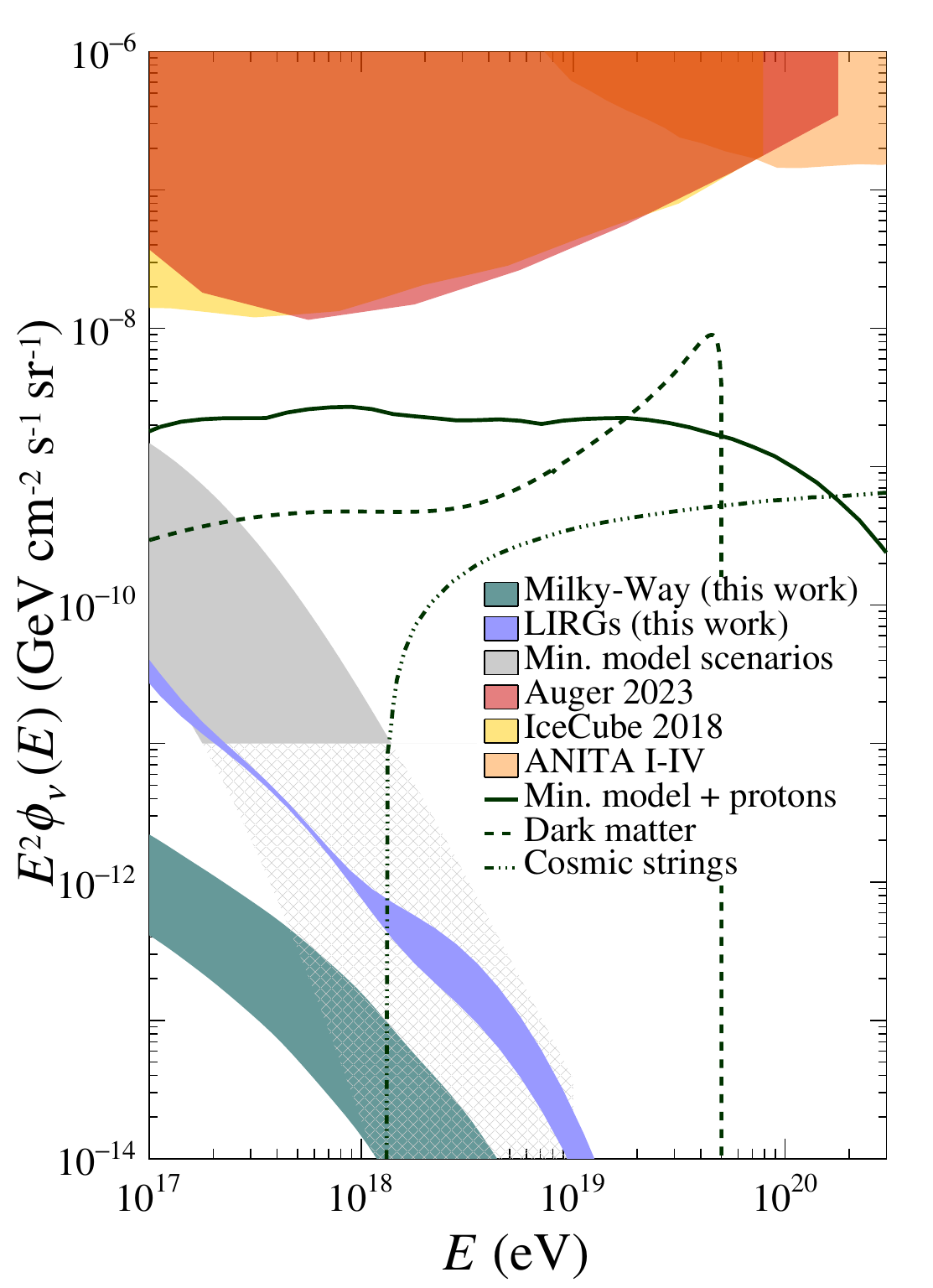}
\caption{Energy flux of neutrinos (single flavor) expected from cosmic-ray interactions in the milky way (``Milky Way'') compared to those from the minimal model of cosmic rays escaping from LIRGs (``LIRGs'') or from various source environments that fit with the minimal model explaining the Auger data above $10^{18.7}~$eV (``Min. model scenarios''). Upper limits from Auger, IceCube and ANITA are reported on top. Also shown are energy fluxes expectations from a non-minimal model of cosmic rays with protons at UHE (``Min. model + protons''), a superheavy dark-matter scenario (``Dark matter'') and a phase transition one (``Cosmic strings'').}
\label{fig:summary}
\end{figure}

In this paper, we have presented an end-to-end calculation of the cosmogenic neutrino flux expected from UHE cosmic ray interactions with the gas in the Milky Way. The main uncertainties stem from the modeling of the gas distribution and from the experimental determination of the mass composition of UHE cosmic rays on Earth. The result is independent, on the other hand, of the various mechanisms governing the production and interaction of cosmic rays; for this reason it can be considered as a floor of cosmogenic neutrino fluxes above $10^{17}~$eV. We have also presented generic estimates of cosmogenic neutrino flux expected from extragalactic production, based on a minimal model of cosmic-ray production to explain the mass-discriminated energy spectra observed on Earth above $5{\times}10^{18}~$eV. These estimates are generally larger than the guaranteed floor aforementioned, in agreement with other estimates derived for more specific environments in the literature.

Our results are summarized in Fig.~\ref{fig:summary}. Upper limits currently obtained with  the Pierre Auger Observatory~\citep{PierreAuger:2023pjg}, the Ice Cube Observatory~\citep{IceCube:2018fhm} and ANITA~\citep{ANITA:2019wyx} are shown as the various red regions. Above $10^{17}~$eV, prospects for neutrino detection based on the minimal model of UHE cosmic-ray production are maximized. However, above $10^{18}~$eV, they appear to be rather dim, if not impossible even for an increase of exposure by one or two orders of magnitude with eventual future detectors. This contrasts with the expectations reported in the early literature mentioned in the introduction or with more recent models based on a mass composition of cosmic rays dominated by protons at UHE~\citep[e.g.][]{Fang:2013vla,Fang:2017tla,Decoene:2019eux}. \\

The detection of UHE neutrinos with current exposures or future ones with observatories such as IceCube-Gen2~\citep{IceCube-Gen2:2020qha}, ARIANNA~\citep{ARIANNA:2019scz}, GRAND~\citep{GRAND:2018iaj},  \mbox{POEMMA~\citep{Anchordoqui:2019omw}} or GCOS~\citep{AlvesBatista:2023lqg} may therefore be instrumental in uncovering new phenomena. Three examples of neutrino energy fluxes unexpected from the contemporary minimal model of UHE cosmic rays are shown in Fig.~\ref{fig:summary}; we briefly describe each of them to conclude this paper.

Non-minimal models of UHE cosmic rays postulate the interplay between two source populations, one of them accelerating a sub-dominant population of protons up to, or even above, $10^{20}~$eV. Uncovering such a sub-dominant population, which is still under-constrained with the current sensitivity of the Pierre Auger Observatory, is one major goal of the upgraded version of the Observatory~\citep[e.g.][]{Anastasi:2023eaz,Suomijarvi:2023pvf,Berat:2023ttm}. Even a small fraction of protons translates into a significant increase of the neutrino flux, which can offer prospects for future detection~\citep{Rodrigues:2020pli,PierreAuger:2022atd,Muzio:2023skc}. As an example, we show as the continuous line the maximum neutrino energy flux realizable for proton sources evolving as the star-formation rate, while remaining compatible with constraints on a putative proton sub-component at UHE~\citep{Muzio:2023skc}. 

Dark matter particles could be superheavy, provided their lifetime is much longer than the age of the universe~\citep[e.g.][]{Kachelriess:2018rty,Alcantara:2019sco,Ishiwata:2019aet,Guepin:2021ljb,Berat:2022iea,PierreAugerCollaboration:2022tlw,Das:2023wtk}. Scotogenic neutrinos are expected to emerge from the cascade of the decaying dark-matter candidate. As an example of extension of the Standard Model of particle physics that includes a superheavy and metastable dark-matter particle, we use the model of \cite{Dudas:2020sbq} as a benchmark. The neutrino energy flux emerging from the decay of the particle is shown as the dotted line for a mass of the particle $M_X=10^{20}~$eV and a mixing angle between active and sterile neutrinos that governs the lifetime of the particle $\theta=10^{-10}$; both values are indeed viable given all known constraints on neutrino and photon flux upper limits at UHE and on the effective number of neutrinos inferred from cosmological observations~\citep{PierreAuger:2023vql}.  

Finally, several extensions of the Standard Model of particle physics predict phase transitions in the early universe that may be revealed through the detection of stochastic gravitational waves resulting from bubble collisions~\citep[e.g.][]{Ellis:2019oqb,Ellis:2020nnr} or neutrinos at UHE in the case of topological defects left after the transition~\citep[e.g.][]{Vilenkin:2000jqa}. We show as the dashed-dotted line the expectations from the decay of cosmic-string cusps that would allow for exploring cosmic-string tensions $G\mu$ as low as $10^{-20}$~\citep{Berezinsky:2011cp}, while the current bound obtained from upper limits on gravitational-wave background energy density  derived from the O3 data of the Advanced LIGO and Advanced Virgo detectors is $G\mu\leq 4{\times}10^{-15}$~\citep{LIGOScientific:2021nrg}. 

These three examples of neutrino energy fluxes  show the potential of UHE neutrino detection to uncover the existence of either a completely new physical phenomena, or particle acceleration mechanisms heretofore never seen or imagined.

\acknowledgments 

We gratefully acknowledge funding from ANR via the grant MultI-messenger probe of Cosmic Ray Origins (MICRO), ANR-20-CE92-0052.  This work was also made possible with the support of the Institut Pascal at Universit\'e Paris-Saclay during the Paris-Saclay Astroparticle Symposium 2022, with the support of the P2IO Laboratory of Excellence (program ``Investissements d’avenir'' ANR-11-IDEX-0003-01 Paris-Saclay and ANR-10-LABX-0038), the P2I axis of the Graduate School Physics of Universit\'e Paris-Saclay, as well as IJCLab, CEA, IPhT, APPEC, the IN2P3 master projet UCMN and EuCAPT ANR-11-IDEX-0003-01 Paris-Saclay and ANR-10-LABX-0038).



\bibliographystyle{aasjournal.bst}
\bibliography{bibliography}

\end{document}